\documentclass[
superscriptaddress,
final,
reprint,
showkeys,
pre,
aps,
floatfix,
]{revtex4-2}

\usepackage[dvips]{graphicx}
\usepackage[utf8]{inputenc}
\usepackage{float}
\usepackage{nameref}
\usepackage{amsmath,amssymb,booktabs,latexsym,MnSymbol,bm}
\usepackage{blkarray}
\usepackage[table]{xcolor}
\usepackage{xcolor}
\usepackage[%
  colorlinks=true,
  urlcolor=blue,
  linkcolor=blue,
  citecolor=blue
]{hyperref}

\begin{document}

\title{Determining liquid crystal properties with ordinal networks and machine learning}

\author{Arthur A. B. Pessa} 
\affiliation{Departamento de F\'isica, Universidade Estadual de Maring\'a - Maring\'a, PR 87020-900, Brazil}

\author{Rafael S. Zola} 
\affiliation{Departamento de F\'isica, Universidade Estadual de Maring\'a - Maring\'a, PR 87020-900, Brazil}
\affiliation{Departamento de F\'isica, Universidade Tecnol\'ogica Federal do Paran\'a, Apucarana, PR 86812-460, Brazil}

\author{Matja{\v z} Perc} 
\affiliation{Faculty of Natural Sciences and Mathematics, University of Maribor, Koro{\v s}ka cesta 160, 2000 Maribor, Slovenia}
\affiliation{Department of Medical Research, China Medical University Hospital, China Medical University, Taichung, Taiwan}
\affiliation{Alma Mater Europaea, Slovenska ulica 17, 2000 Maribor, Slovenia}
\affiliation{Complexity Science Hub Vienna, Josefst{\"a}dterstra{\ss}e 39, 1080 Vienna, Austria}

\author{Haroldo V. Ribeiro} 
\email{hvr@dfi.uem.br}
\affiliation{Departamento de F\'isica, Universidade Estadual de Maring\'a - Maring\'a, PR 87020-900, Brazil}

\date{\today}

\begin{abstract}
Machine learning methods are becoming increasingly important for the development of materials science. In spite of this, the use of image analysis in the development of these systems is still recent and underexplored, especially in materials often studied via optical imaging techniques such as liquid crystals. Here we apply the recently proposed method of ordinal networks to map optical textures obtained from experimental samples of liquid crystals into complex networks and use this representation jointly with a simple statistical learning algorithm to investigate different physical properties of these materials. Our research demonstrates that ordinal networks formed by only 24 nodes encode crucial information about liquid crystal properties, thus allowing us to train simple machine learning models capable of identifying and classifying mesophase transitions, distinguishing among different doping concentrations used to induce chiral mesophases, and predicting sample temperatures with outstanding accuracy. The precision and scalability of our approach indicate it can be used to probe properties of different materials in situations involving large-scale datasets or real-time monitoring systems.
\end{abstract}

\maketitle

\section{Introduction}

The use of machine learning methods for probing physical and chemical properties of materials is becoming increasingly popular among researchers working in physics~\cite{baldi2014searching, mukund2017transient, carleo2019machine, dreissigacker2019deep}, chemistry~\cite{ma2018deep, zhangprediction2019}, and materials science~\cite{butler2018machine, jha2018elemnet, ziletti2018insightful, wei2019machine, tshitoyan2019unsupervised}. These recent developments are closely related to important advances in computational technologies, novel statistical learning methods, and the availability of large datasets. The combination of these scientific advances has great potential to revolutionize the role of computational methods in applied research and has been considered by many as the `fourth paradigm of science' in materials science~\cite{agrawal2016perspective} capable of paving the way to the fourth industrial revolution.

Several of these works, particularly those involving biological and complex materials, use optical imaging techniques and benefit from their non-destructive features. Liquid crystals are a typical example of materials that have been, and still are, extensively studied through imaging methods. This is because liquid crystals are birefringent, so that polarized optical microscope imaging often suffices to determine many different properties of these materials~\cite{degennes1993thephysics, lee2016optical}. Despite that, the use of machine learning methods in liquid crystals remains surprisingly limited, and indeed only a few works have tried to directly associate physical properties of these materials with their optical textures~\cite{sigaki2019estimating, sigaki2020learning}. As in other machine learning problems (regressions or classifications) involving images, one can learn the underlying physics of liquid crystals by extracting features from optical textures and training algorithms with a set of examples consisting of images and their associated physical properties. The process of extracting features from images to be used in classification or regression problems often demands domain knowledge about the material structure~\cite{ramprasad2017machine, butler2018machine, murdock2020isdomain}, and it is often helpful to have more general methods capable of generating image features regardless of the particularities of a material. 

An interesting possibility is to use image features derived from permutation entropy and the ordinal symbolization introduced by Bandt and Pompe~\cite{bandt2002permutation}. Initially proposed for time series and later generalized to higher-dimensional data structures~\cite{ribeiro2012complexity}, the Bandt-Pompe approach consists of estimating a probability distribution associated with the occurrence of ordinal patterns at very local scales of a dataset (typically 2$\times$2 pixels in images and usually less than seven elements in time series). The Shannon entropy of this ordinal distribution defines the permutation entropy that, together with statistical complexity~\cite{lopezruiz1995statistical, rosso2007distinguishing}, are the most commonly used quantifiers associated with the Bandt-Pompe symbolization method~\cite{zanin2021ordinal}. Beyond counting ordinal patterns, recent developments related to the Bandt-Pompe approach have shown that analyzing transitions among ordinal patterns as complex networks can be even more effective for characterizing datasets~\cite{small2013complex, mccullough2017multiscale, small2018ordinal, pessa2019characterizing, pessa2020mapping}. These ordinal networks were also originally proposed for mapping time series into complex networks~\cite{small2013complex} and only very recently have been extended to produce a map between images and complex networks~\cite{pessa2020mapping}.

Due to their recency, the use of ordinal methods --- particularly ordinal networks --- to characterize images of physical systems remains limited. Here, we reduce this shortage by combining image features extracted from ordinal networks and a simple machine learning method to investigate experimental samples of liquid crystals. Our research shows that the combination of these tools is quite effective in identifying phase transitions, distinguishing between first-order and second-order transitions, and predicting doping concentrations and sample temperatures. The impressive accuracy of learning methods trained with features directly extracted from ordinal networks allows us to conjecture that this approach is likely to be very helpful to solve other problems in materials science that involve imaging techniques.

\section{Experimental procedures and dataset of textures}\label{sec:exp}

We start by describing our experiments with liquid crystal samples used to obtain a dataset of textures of these materials (data and code are available at \url{gitlab.com/arthurpessa/ordinal_nets_lc}). These experiments were carried out with 3 samples of 8CB and 24 samples of E7 liquid crystal hosts mixed with one or two chiral dopants (all purchased from Merck) at different weight percentages and other 2 samples of 8CB and E7 without dopants. The 3 doped 8CB samples were mixed with 3.26\% of the R811 dopant (right-handed enantiomer). The 26 E7 samples were divided into two main groups. The first group consists of 2 pure samples and 10 samples mixed with the R811 dopant at different concentrations ranging from 0.75\% to 23.20\%. The second group comprises 16 samples mixed with the R811 and the S811 (left-handed enantiomer) dopants, such that the concentrations of the R811 are always smaller than the S811 (allowing the pitch to vary), but the amount of doping remained fixed at 10\% of total sample weight (here called racemic mixtures). Table~\ref{tab:1} shows detailed information about the doping concentrations in each sample. At room temperature, the pure 8CB samples display a smectic A mesophase and form a chiral smectic A mesophase after being doped with R811. In turn, the pure E7 samples present a nematic mesophase and become cholesteric once doped with the chiral additive. The different concentrations of dopants translate into different cholesteric pitches. The samples with fixed amount additives allow the cholesteric pitch to change without changing other physical properties of the mesophase. 

\begin{table}[!t]
    \centering
    \caption{Summary of the liquid crystal samples used in our study. The percentages of chiral dopants (R811 and S811) are relative to sample weight, and the number of images refer to all samples with the same doping concentration.}\label{tab:1}
    \begin{tabular}{lrrcc}
    \hline
    Host  &  R811    &  S811    & Samples & Images \\ 
    \hline 
    8CB   &  0.00\%  &  0.00\%  & 2       & 323    \\
    8CB   &  3.26\%  &  0.00\%  & 3       & 495    \\ [.25em]
    E7    &  0.00\%  &  0.00\%  & 2       & 316    \\
    E7    &  0.75\%  &  0.00\%  & 2       & 316    \\
    E7    &  3.14\%  &  0.00\%  & 2       & 309    \\
    E7    &  5.50\%  &  0.00\%  & 2       & 301    \\
    E7    &  8.00\%  &  0.00\%  & 2       & 282    \\
    E7    & 23.20\%  &  0.00\%  & 2       & 186    \\ [.25em]
    E7    &  0.00\%  & 10.00\%  & 3       & 451    \\ 
    E7    &  1.00\%  &  9.00\%  & 3       & 470    \\
    E7    &  2.00\%  &  8.00\%  & 3       & 473    \\
    E7    &  3.00\%  &  7.00\%  & 1       & 150    \\
    E7    &  4.00\%  &  6.00\%  & 2       & 331    \\ 
    E7    &  4.85\%  &  5.15\%  & 2       & 331    \\ 
    \hline
    \end{tabular}
\end{table}

\begin{figure*}[!t]
    \centering
    \includegraphics[width=1\linewidth]{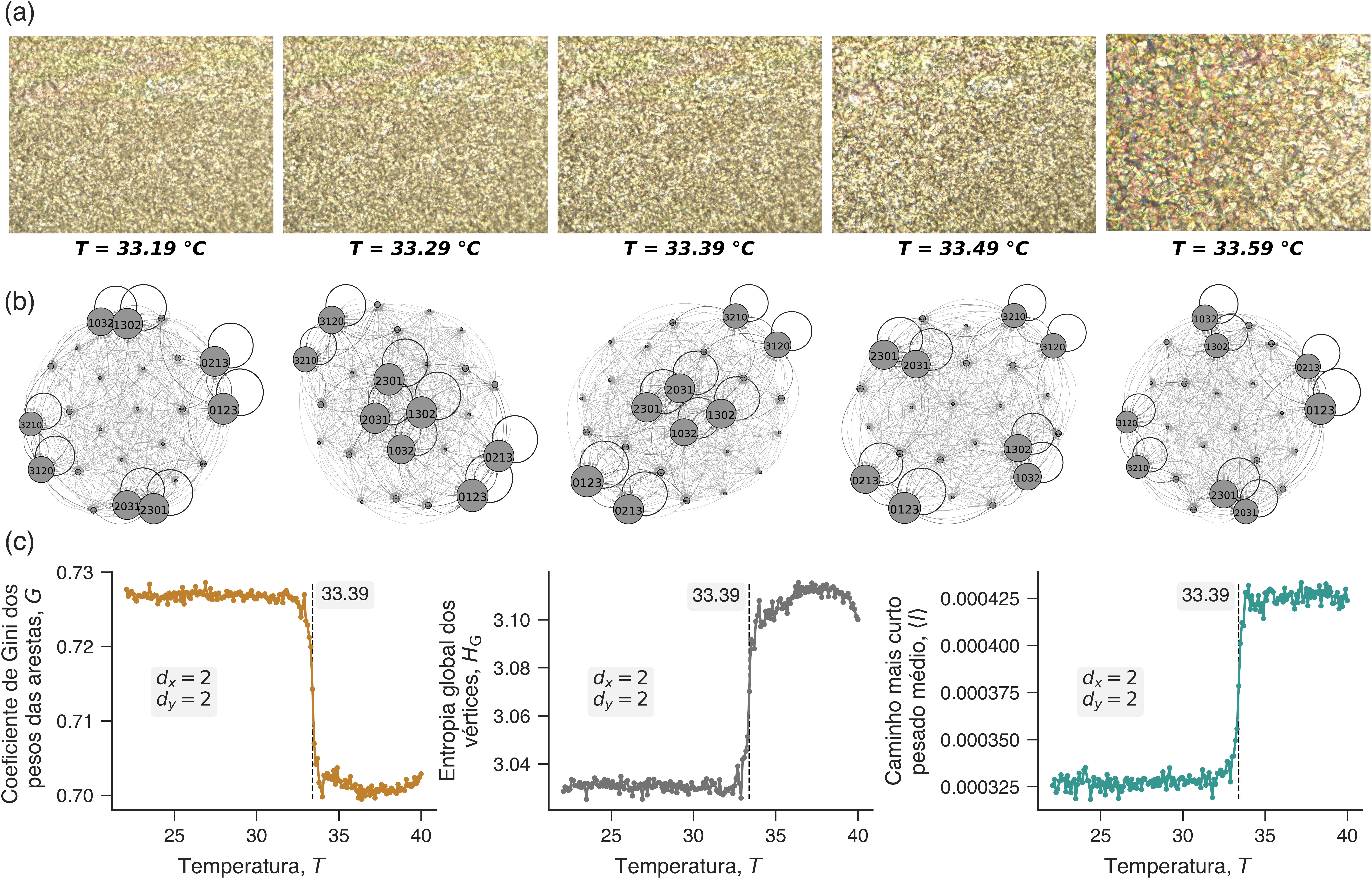}
    \caption{Identifying mesophase transitions in liquid crystal samples with ordinal networks. (a) Examples of optical textures obtained from an 8CB liquid crystal sample doped with 3.26\% of R811 at 5 temperatures around the critical temperature $T_c = 33.39^{\circ}{}\rm{C}$ associated with a second-order transition between the mesophases chiral smectic A and cholesteric. (b) Visualization of the ordinal networks mapped from the images in the previous panel with $d_x = d_y = 2$. We have made node sizes proportional to the occurring frequency of their associated ordinal patterns and used a grayscale color map to highlight edge weights (the darker the shade, the higher the weight). (c) Dependence with the temperature of three network metrics calculated from ordinal networks mapped from the same 8CB liquid crystal sample doped with 3.26\% of R811. We observe that the three metrics undergo abrupt changes precisely at the critical temperature (indicated by dashed lines).}
\label{fig:1}
\end{figure*}

To obtain typical textures of each mesophase and avoid flow alignment, all samples are capillary filled into hollow rectangular borosilicate glass capillaries (100$\mu$m $\times$ 1.0mm) without internal treatment at $60^\circ$C. Next, the samples are cooled up to room temperature and placed on a temperature controller (Instec MK2000) under a polarized light microscope setup (Leica). We then heat these samples at $0.1^\circ$C/min in temperature ranges involving first-order and second-order mesophase transitions while taking pictures of the textures every 10s with a camera (Leica ICC50W) attached to the microscope. After reaching a maximum temperature, we cool down the samples at $-0.1^\circ$C/min while taking pictures of the textures every 10s. After cropping, these images have 800 $\times$ 1070 pixels of size and are saved in JPEG format with 24 bits per pixel (8 bits for red, green, and blue colors in the RGB color space). The 8CB samples (pure or doped) first undergo a second-order transition (smectic A to nematic for pure and chiral smectic A to cholesteric for doped samples) before presenting a first-order transition (nematic to isotropic for pure and cholesteric to isotropic for doped samples), whereas the E7 samples display a first-order transition (nematic to isotropic for pure and cholesteric to isotropic for doped samples). Figure~\ref{fig:1}{a} shows examples of textures obtained from an 8CB sample doped with 3.26\% of R811 in a temperature range where the second-order transition takes place. We observe that textures immediately before and after the phase transition are very similar, such that even a well-trained eye is likely to have great difficulty in precisely identifying the critical temperature only by visually inspecting these images. Textures obtained from E7 or 8CB samples immediately before the first-order transitions are equally hard to distinguish visually. 

\section{Ordinal networks}\label{sec:methods}

After obtaining the typical textures from the mesophases present in our samples, we average the image pixels over the three RGB layers to represent each texture as a simple array $M=\{y_t^u\}^{u = 1, \dots, N_y}_{t = 1, \dots, N_x}$, where $y_t^u$ represents the average pixel intensity at column $t$ and line $u$, while $N_x$ and $N_y$ are respectively the image width and height. We then map these two-dimensional arrays into ordinal networks~\cite{pessa2020mapping, pessa2021ordpy}. To illustrate this method, suppose we have a $3\times3$ pixels image represented by
\begin{equation}\label{eq:data_array}
    M = 
    \begin{pmatrix}
        4 & 8 & 3\\
        6 & 7 & 5\\
        2 & 8 & 9\\
    \end{pmatrix}\,.
\end{equation}
The first step of the algorithm is to partition the matrix $M$ into overlapping submatrices of size $d_x \times d_y$ (the embedding dimensions). While other choices are certainly possible~\cite{ribeiro2012complexity, zunino2016discriminating, pessa2020mapping}, it is common to set $d_x=d_y=2$ when dealing with images~\cite{pessa2020mapping}. For this choice, the matrix $M$ is partitioned into four submatrices $\{w_p^q\}_{p=1,\dots,n_x}^{q=1,\dots,n_y}$ (with $n_x=N_x-d_x+1$ and $n_y=N_y-d_y+1$): $w_1^1 = \begin{pmatrix} 4 & 8\\ 6 & 7\\ \end{pmatrix}$, $w_2^1 = \begin{pmatrix} 8 & 3\\ 7 & 5\\ \end{pmatrix}$, $w_1^2 = \begin{pmatrix} 6 & 7\\ 2 & 8\\ \end{pmatrix}$, and $w_2^2 = \begin{pmatrix} 7 & 5\\ 8 & 9\\ \end{pmatrix}$. These matrices are then horizontally flattened, yielding: $w_1^1 = (4,8,6,7)$, $w_2^1 = (8,3,7,5)$, $w_1^2 = (6,7,2,8)$, and $w_2^2 = (7,5,8,9)$. The core of the ordinal network algorithm is to apply the Bandt-Pompe symbolization approach~\cite{bandt2002permutation} in order to obtain another array $\{\pi_p^q\}_{p=1,\dots,n_x}^{q=1,\dots,n_y}$ representing the ordinal pattern $\pi_p^q$ associated with each data partition. These ordinal patterns are obtained by evaluating the permutation of the index numbers $(0,1,\dots,d_x d_y-1)$ that sorts the elements of $w_p^q$ in ascending order. For instance, the sorted version of $w_1^1$ is $(4,6,7,8)$ and so this partition is associated with the ordinal symbol $\pi_1^1=(0,2,3,1)$, where the index number $0$ corresponds to the position of the number $4$ in $w_1^1$, $2$ corresponds to the position of the number $6$ in $w_1^1$, and so on. Similarly, we find that $w_2^1$, $w_1^2$, and $w_2^2$ are respectively associated with $\pi_2^1=(1,3,2,0)$, $\pi_1^2=(2,0,1,3)$, and $\pi_2^2=(1,0,2,3)$.

Having obtained the array of permutation symbols $\{\pi_p^q\}$, we consider all its unique permutations $\Pi_i$ $ [i=1,\dots,(d_x d_y)!]$ as nodes of the ordinal network. These nodes are connected by directed edges if they appear vertically ($\pi_p^q \to \pi_p^{q+1}$ for $q = 1,\dots,n_y-1$) or horizontally ($\pi_p^q \to \pi_{p+1}^{q}$ for $p = 1,\dots,n_x-1$) adjacent in the symbolic array. These links are further weighted according to the relative occurrence of each transition in the symbolic array, such that we can write the elements $\rho_{i,j}$ of the weighted adjacency matrix $A$ representing the ordinal network as
\begin{equation}\label{eq:edge_weights}
    \rho_{i,j} = \frac{\text{total of transitions $\Pi_i \to \Pi_j$ in $\{\pi_p^q\}$}}{2n_x n_y-n_x-n_y}\,,
\end{equation}
with $i, j = 1, 2, \dots$ standing for all permutations $\Pi_i$ occurring in $\{\pi_p^q\}$. Following this procedure, we can map our $3\times3$ example image (Eq.~\ref{eq:data_array}) into an ordinal network composed by four nodes [$\Pi_1 = (0,2,3,1)$, $\Pi_2 = (1,0,2,3)$, $\Pi_3 = (1,3,2,0)$ and $\Pi_4 = (2,0,1,3)$] and represented the following adjacency matrix:
\begin{equation}
A = \!\!
\renewcommand\arraystretch{1.2}
\begin{blockarray}{l *{4}{c}}
    \begin{block}{l *{4}{>{$\footnotesize}c<{$}}}
      & 0231 & 1023 & 1320 & 2013 \\
    \end{block}
    \begin{block}{>{$\footnotesize}l<{$} [*{4}{c}]}
    0231 & 0 & 0 & 1/4 & 1/4\\
    1023 & 0 & 0 & 0 & 0\\
    1320 & 0 & 1/4 & 0 & 0\\
    2013 & 0 & 1/4 & 0 & 0\\
    \end{block}
  \end{blockarray}\;.
\end{equation}
Figure~\ref{fig:1}{b} depicts visualizations of ordinal networks mapped from textures collected around the transition between the mesophases chiral smectic A and cholesteric of an 8CB sample doped with 3.26\% of R811. We have used $d_x=d_y=2$ in these examples and in all other results of this work, which in turn constrains our ordinal networks to have up to 24 nodes and 416 edges~\cite{pessa2020mapping}. Furthermore, we have used the numerical implementation of ordinal networks available in the Python module \texttt{ordpy}~\cite{pessa2021ordpy}.

After mapping images into ordinal networks, we can use standard network metrics~\cite{newman2010networks} to characterize our liquid crystal textures. In addition to usual network metrics, ordinal networks have also two entropy-related quantifiers~\cite{mccullough2017multiscale, small2018ordinal, pessa2019characterizing, pessa2020mapping, sakellariou2019markov}. At the vertex level, we can calculate the local node entropy for node $i$ (permutation $\Pi_i$) as $s_i = -\sum_{j\in\mathcal{O}_i}\rho'_{i, j}\log \rho'_{i,j}$, where $\rho'_{i, j} = \rho_{i, j}/\sum_{k \in\mathcal{O}_i}\rho_{i, k}$ represents the renormalized probability of transitioning from permutation nodes $i$ to $j$ (permutation symbols $\Pi_i$ and $\Pi_j$), and $\mathcal{O}_i$ is the outgoing neighborhood of node $i$. This quantity measures the determinism of transitions among permutation symbols at the node level, such that $s_i$ is maximum when all edge weights of edges stemming from $i$ are uniformly distributed, while $s_i=0$ solely if one edge leaves the node $i$. Using the values of $s_i$, we can further evaluate the global node entropy~\cite{mccullough2017multiscale, small2018ordinal, pessa2020mapping} as
\begin{equation}\label{eq:global_node_entropy}
    S_{\rm GN} = \sum_{i=1}^{n_\pi} \rho_i s_i\,,
\end{equation}
where $n_\pi$ is the number of nodes in the ordinal network and $\rho_i$ is the probability of finding the permutation symbol $\Pi_i$ in the symbolic array.

\section{Results}

In an initial test with ordinal networks mapped from liquid crystal textures, we investigate whether simple network metrics can efficiently detect the different mesophase transitions displayed by our samples. To do so, we map all textures in our dataset into ordinal networks with $d_x = d_y = 2$ and evaluate the temperature dependence of three network metrics for all samples. Following Ref.~\cite{pessa2020mapping}, we calculate the Gini coefficient $G$ associated with edge weights of ordinal networks, the global node entropy $S_{\rm GN}$ (Eq.~\ref{eq:global_node_entropy}), and the average weighted shortest path $\langle l \rangle$. Figure~\ref{fig:1}{c} illustrates the dependence of these three quantities on the temperature $T$ for an 8CB sample doped with 3.26\% of R811. We observe that the three network metrics show abrupt variations at $T=33.39^{\circ}{}\rm{C}$, a value that agrees with the critical temperature related to the transition between the mesophases chiral smectic A and cholesteric reported in the literature~\cite{lin2015phase}. These marked and abrupt changes occur around the phase transition temperatures of all samples. To systematically estimate these critical temperatures, we determine the temperature associated with the largest change in each network metric and use a majority vote strategy to select a single temperature value to represent our estimate for the critical temperature. We find this strategy to work quite well, yielding estimates that almost perfectly agree with values reported in the literature~\cite{zola2013surface, lin2015phase}.

In another application related to mesophase transitions, we ask whether the information encoded in ordinal networks can discriminate between first-order and second-order transitions. To investigate this possibility, we set up a binary classification task in which liquid crystal textures near the critical temperature are used to train a $k$-nearest neighbors classifier --- a very simple and intuitive learning algorithm~\cite{pedregosa2011scikitlearn, james2014introduction} --- to decide whether the associated transition is of first or second order (all algorithms used in this work were implemented using the Python module \texttt{scikit-learn}~\cite{pedregosa2011scikitlearn}). The weights of all 416 possible edges of ordinal networks with $d_x=d_y=2$ are used as features in this classification task (edges not occurring are assigned zero weight). We select five images for each sample in our study (two before, two after, and one at the critical temperature) and their corresponding mesophase transitions. It is worth remembering that our samples undergo four different transitions, two of first order (nematic to isotropic and cholesteric to isotropic) and two of second order (smectic A to nematic and chiral smectic A to cholesteric). 

\begin{figure*}[!t]
    \centering
    \includegraphics[width=0.7\linewidth]{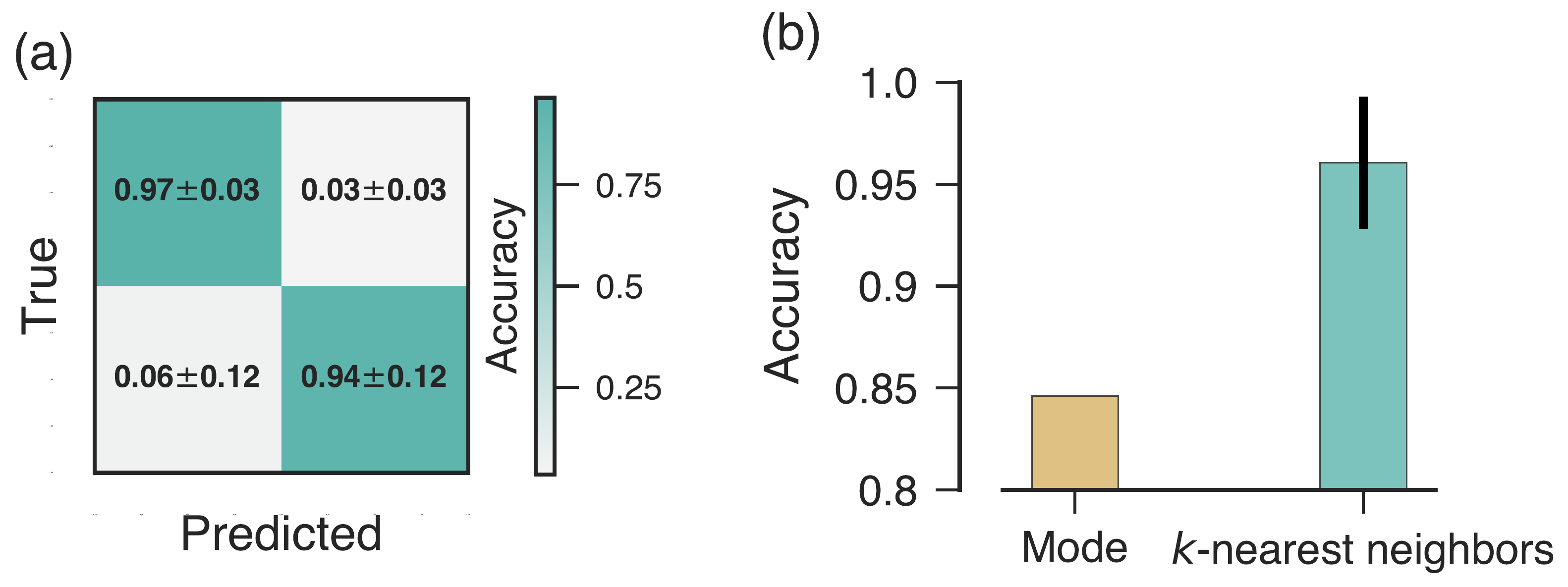}
    \caption{Determining the order of mesophase transitions with ordinal networks. (a) Confusion matrix associated with the task of distinguishing between first-order and second-order transitions. The values are an average over 100 $k$-nearest neighbor classifiers trained with different train and test sets randomly split ($\pm$ indicates one standard deviation). (b) Average accuracy of the $k$-nearest neighbors method ($\approx$96\%) compared with the accuracy of a dummy classifier ($\approx$85\%) that always predicts the mode of transition orders (the error bar stands for one standard deviation).}
\label{fig:2}
\end{figure*}

This procedure leads us to a dataset comprising 155 images mapped into ordinal networks, 130 related to first-order transitions and 25 associated with second-order transitions. We separate these data into training ($75$\%) and test ($25$\%) sets (stratifying by the different classes) and perform a grid search (with 5-fold cross-validation strategy) on the training set to optimize the number of nearest neighbors (the only parameter of the algorithm) in the model~\cite{pedregosa2011scikitlearn, james2014introduction}. Because of the relatively small number of images, we further generate 100 random partitions of training and test sets to estimate the average performance of the best classification models. Figure~\ref{fig:2}{a} depicts the confusion matrix associated with this binary classification problem, where we observe that ordinal networks are very good at identifying the order of the mesophase transitions. Indeed, Figure~\ref{fig:2}{b} shows that the average performance of the $k$-nearest neighbors classifiers is around 96\%, significantly outperforming the accuracy of a dummy classifier that assigns the most common class in the training set (first-order transition) to all instances in the test set.

We now focus on the two groups of E7 samples (Table~\ref{tab:1}) to investigate whether ordinal networks are capable of discriminating among different doping concentrations regardless of the sample temperature. All these samples present transitions to the isotropic phase at high temperatures, and are thus indistinguishable when reaching this mesophase (isotropic textures are entirely black). Because of this, we restrict our analysis to textures obtained up to  $1^{\circ}{}\rm{C}$ before the critical temperature. It is worth noticing that the critical temperatures of E7 samples doped only with R811 decrease as the doping concentration increases, such that the temperature ranges vary among samples with different doping concentrations. Conversely, the E7 samples with a fixed amount of doping (racemic mixtures of R811 and S111) display practically the same critical temperatures and are constrained to approximately the same temperature range in this analysis. Thus, discriminating among racemic samples can be considered (at least in principle) more challenging, since the different temperature ranges of E7 samples doped only with R811 add additional information to the classification tasks that may eventually help to distinguish these samples.

\begin{figure*}[!t]
    \centering
    \includegraphics[width=0.7\linewidth]{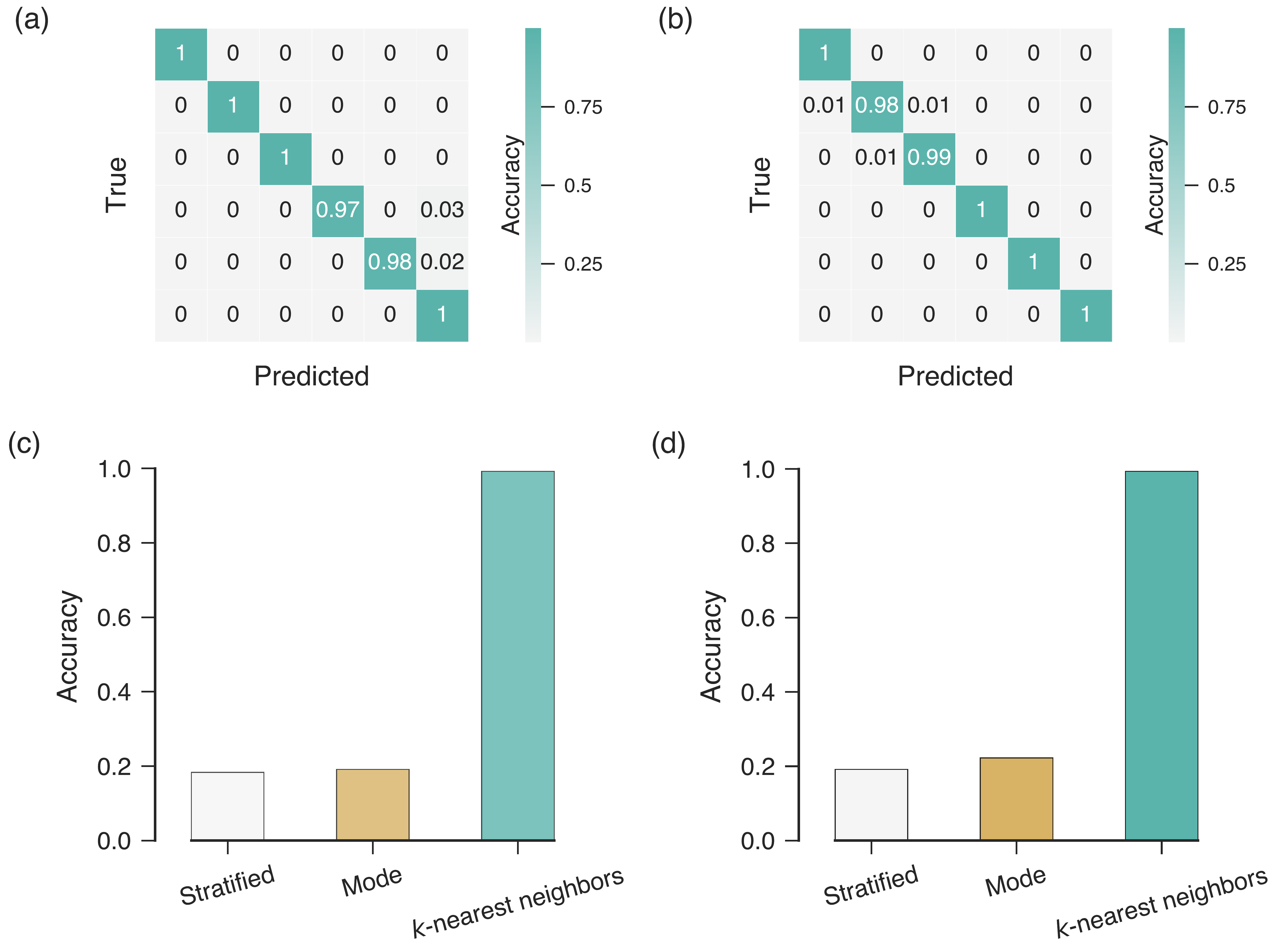}
    \caption{Distinguishing among different doping concentrations with ordinal networks. (a) Confusion matrix associated with the predictions of the trained $k$-nearest neighbors classifier on the test set of E7 samples doped only with R811. The first matrix row corresponds to 0\% of R811, the second to 0.75\% of R811, and so on, following the same order of Table~\ref{tab:1}. (b) The analogous confusion matrix obtained for E7 samples doped with racemic mixtures of R811 and S811. In this case, the first matrix row corresponds to 0\% of R811 and 10\% of S811, the second to 1\% of R811 and 9\% of S811, and so on, following the same order of Table~\ref{tab:1}. (c) Accuracy of the trained $k$-nearest neighbors classifier for the E7 samples doped only with R811 compared to dummy classifiers that make predictions based on the relative frequency of each doping concentration (stratified) and on the most frequent doping concentration (mode). (d) The same as in the previous panel, but estimated for E7 samples doped with racemic mixtures of R811 and S811.}
\label{fig:3}
\end{figure*}

Similarly to the classification of transition orders, we use the $k$-nearest neighbors algorithm to classify the doping concentrations of E7 samples. We have a total of 1,527 images for the E7 samples doped only with R811 at six different concentrations and 1,813 images of E7 samples doped with six different racemic mixtures of R811 and S811 (Table~\ref{tab:1}). In both cases, we separate the data into training (75\%) and test (25\%) sets, stratifying by the six different classes of samples (the different doping concentrations). Again, we use all the 416 edge weights of the ordinal networks mapped from these images with $d_x=d_y=2$ as features to train the learning method. We further consider a grid search and a 5-fold cross-validation strategy to tune the number of nearest neighbors in the $k$-nearest neighbors algorithm. Figures~\ref{fig:3}{a} and \ref{fig:3}{b} depict the confusion matrices obtained by applying the trained algorithm to the test sets of both types of samples. We observe that the diagonal elements of these matrices are very close to one, indicating that our approach almost perfectly classifies the doping concentrations. We have also verified that the accuracy of the $k$-nearest neighbors classifiers trained with the weights of ordinal networks significantly outperforms the accuracy of dummy classifiers based on the frequency of each class or the mode of classes in the training set, as shown in Figures~\ref{fig:3}{c} and \ref{fig:3}{d}. Thus, these results imply that ordinal networks are equally good at predicting doping concentrations of samples doped only with R811 or with racemic mixtures of R811 and S811. 

To further demonstrate the versatility of our approach, we propose now to predict the sample temperatures independently of doping concentration. Once again, we consider, separately, E7 samples doped only with R811 and E7 samples doped with racemic mixtures of R811 and S811. This task represents a regression problem in which we train a learning algorithm to predict the sample temperature using the weights of ordinal networks. For the E7 samples doped only with R811, we select images in the temperature range between $25^{\circ}{}\rm{C}$ and $44^{\circ}{}\rm{C}$ to ensure that each sample does not undergo a phase transition. In the case of E7 samples doped with racemic mixtures of R811 and S811, the temperature range is homogeneous and allows us to consider textures within a range of temperatures from $41^{\circ}{}\rm{C}$ to $54^{\circ}{}\rm{C}$. This selection leads us to 898 images for E7 samples doped only with R811 and 1,454 textures related to racemic mixtures of R811 and S811. To increase image sampling over the temperature ranges, we slice each image into six non-overlapping parts of size $800 \times 175$ pixels before mapping each one into ordinal networks with $d_x=d_y=2$. 

\begin{figure*}[!t]
    \centering
    \includegraphics[width=0.7\linewidth]{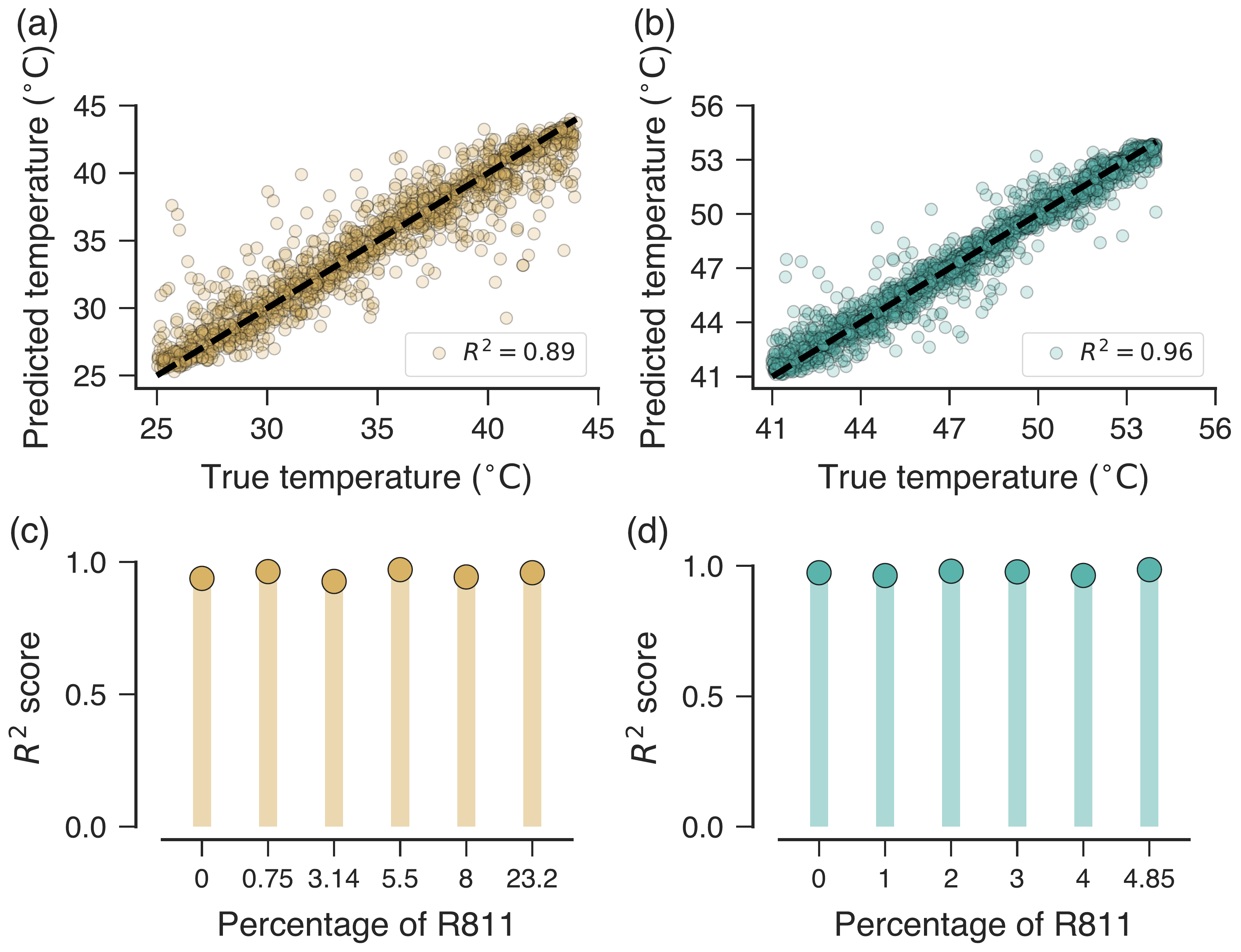}
    \caption{Predicting the temperature of E7 liquid crystal samples with ordinal networks. (a) Relationship between the predicted and true sample temperatures obtained by applying the trained $k$-nearest neighbors model to the test set comprising E7 samples doped only with R811. (b) The same relationship as in the previous panel obtained for E7 samples doped with racemic mixtures of R811 and S811. (c) Coefficients of determination obtained from $k$-nearest neighbors models trained with ordinal network weights to predict the temperature of E7 samples doped only with R811 after grouping the data by each concentration of R811. (d) The same as in the previous panel, but determined for each combination of the E7 racemic mixtures of R811 and S811.}
\label{fig:4}
\end{figure*}

We have again separated the datasets representing both types of samples into training (75\%) and test (25\%) sets and performed a $k$-nearest neighbors regression with the number of nearest neighbors optimized using a grid search in a 5-fold cross-validation strategy. Figures~\ref{fig:4}{a} and \ref{fig:4}{b} show the relationships between the true and predicted temperatures for the E7 samples doped only with R811 and for the racemic mixtures of R811 and S811. We observe the data points are concentrated near the 1:1 relation. Indeed, the coefficients of determination ($R^2$ scores) are $R^2 = 0.89$ for E7 samples doped only with R811 and $R^2 = 0.96$ for E7 samples doped with racemic mixtures of R811 and S811, reinforcing the good precision of our method.

In a last application, we propose to predict the sample temperatures after grouping the data by each concentration of R811 and by each combination of the racemic mixtures of R811 and S811 (Table~\ref{tab:1}). In this case, we use the largest temperature range available in each group of samples. Again, we slice all images into 6 non-overlapping parts and map each one into ordinal networks with $d_x=d_y=2$. These datasets are then split into training (75\%) and test (25\%) sets. For each group of samples, we train a $k$-nearest neighbors algorithm using the weights of ordinal networks to predict the sample temperatures (with number of nearest neighbors optimized via a grid search in a 5-fold cross-validation strategy). Figures~\ref{fig:4}{c} and \ref{fig:4}{d} show that $R^2$ scores are always larger than $0.92$ for all types of E7 samples and even better ($R^2 > 0.96$) for the group of samples with racemic mixtures of R811 and S811.

Taken together, the outstanding performance obtained in our applications with liquid crystal samples and simple machine learning methods indicates that ordinal networks are an excellent representation of liquid crystal textures capable of encoding information about their mesophases, doping concentrations, and temperatures.

\section{Conclusions}

We have presented a comprehensive investigation of liquid crystal properties using optical textures obtained from several experimental samples. We have mapped these images into ordinal networks formed by only 24 nodes and asked whether this representation can encode essential information about liquid crystals. Our research has addressed this question by combining ordinal networks with a simple machine learning method (the $k$-nearest neighbors algorithm) to predict different physical properties of liquid crystals. This approach has shown to be quite effective in identifying and classifying mesophase transitions, distinguishing among different doping concentrations, and determining sample temperatures. Furthermore, beyond the impressive accuracy obtained in all classification and regression tasks, ordinal networks naturally inherit all advantages of the Bandt-Pompe method~\cite{bandt2002permutation}, which are computational efficiency, robustness against noisy data, and invariance under monotone increasing nonlinear scaling of data. This in turn makes our approach easily scalable to situations involving large-scale datasets or real-time monitoring systems. These features corroborate ordinal networks as a simple and versatile tool for extracting image features that are likely to be useful in other problems of materials science. Thus, we hope our research motivates other investigations that use ordinal networks to probe physical properties of different materials.

\acknowledgments
We acknowledge the support of the Coordena\c{c}\~ao de Aperfei\c{c}oamento de Pessoal de N\'ivel Superior (CAPES), the Conselho Nacional de Desenvolvimento Cient\'ifico e Tecnol\'ogico (CNPq -- Grants 407690/2018-2, 303121/2018-1 and 304634/2020-4), the LAMAP-UTFPR, and the Slovenian Research Agency (Grants J1-2457 and P1-0403).

\bibliographystyle{elsarticle-num.bst}

\end{document}